\documentclass[10pt.]{article}

\usepackage{amsmath}
\usepackage{amssymb}

\usepackage{graphicx}

\usepackage{cite}
\usepackage{color}

\makeatletter
\renewcommand{\@biblabel}[1]{\quad#1.}
\makeatother

\date{}

\begin{document}
\begin{flushleft}

{\Large
\textbf{The emergence of pseudo-stable states in network dynamics}
}\\
L. Hedayatifar$^{1}$, F. Hassanibesheli$^1$, A.~H. Shirazi$^1$, S. Vasheghani Farahani $^2$, G.~R. Jafari$^{1,3,\ast}$
\\
\bf{1} Department of Physics, Shahid Beheshti University, G.C., Evin, Tehran 19839, Iran\\
\bf{2} Department of Physics, Tafresh University, Tafresh 39518 79611, Iran\\
\bf{3} The Institute for Brain and Cognitive Science (IBCS), Shahid Beheshti University,\\
 G.C., Evin, Tehran 19839, Iran\\
\bf{4} Center for Network Science, Central European University, H-1051, Budapest, Hungary\\
$\ast$ E-mail: g\_jafari@sbu.ac.ir
\end{flushleft}

\date{\today}

\section*{Abstract}

In the context of network dynamics, the complexity of systems
increases possible evolutionary paths that often are not
deterministic. Occasionally, some map routs form over the course of
time which guide systems towards some particular states. The main
intention of this study is to discover an indicator that can help predict
these pseudo-deterministic paths in advance. Here we
investigate the dynamics of networks based on Heider balance theory
that states the tendency of systems towards decreasing tension. This
inclination leads systems to some local and global minimum tension
states called "jammed states" and "balanced states", respectively.
We show that not only paths towards jammed states are not completely
random but also there exist secret pseudo deterministic paths that
bound the system to end up in these special states. Our results
display that the Inverse Participation Ratio method (IPR) can be a
suitable indicator that exhibits collective behaviors of systems.
According to this method, these specific paths are those that host
the most participation of the constituents in the system. A direct
proportionality exists between the distance and the selectable paths
towards local minimums; where by getting close to the final steps there is no other way but
the one to the jammed states.
\textbf{keywords:} Network dynamics, Balance theory, Jammed state, Participation\\\\
\section*{Introduction}

In every society there is a common question; what happens tomorrow,
or what comes next? It is not easy to answer this question unless
you gain insight on the A to Z of that society. A society consists
of a collection of members that are not exactly independent of each
other. In fact they might very much influence each other either by
gossip \cite{Traag} or in a viral fashion \cite{Kaj}. This
dependency increases the complexity and possible paths that a system
can experience. As a self-organized process, societies evolve
towards a lower stress status among members (the least stress
principle) \cite{heider,car,antal}. This principle deforms the
topology of societies, eventually gaining stability
\cite{marvel}. Now how could one gain a physical insight on the
structure and dynamics of such systems? One answer owes its
existence to the concept of Hamiltonian. As such, a smart technique
that could be implemented to deal with such issues is to relate a
Hamiltonian to the system and study its evolution \cite{axel}.
Although the Hamiltonian equations provide a platform for
issuing statements on the tendency of societies towards lower energy
states \cite{marvel,qian}, but due to the number of possible
paths, the Hamiltonian could not give us a clear picture of the next
steps. The query that we try to answer here is whether there is an
indicator to uncover hidden pseudo-deterministic paths towards
special states? In the present study since we will be dealing with the sociability of the members in a society, we implement the Heider balance theory \cite{heider}. The decrease of energy in
these societies would eventually lead to local or global
minimum energy states named as the jammed states or balanced states
\cite{antal,marvel}.
\begin{figure}
\centerline{\includegraphics[trim = 5mm 1mm 1mm 0mm, clip,width=.5\textwidth]{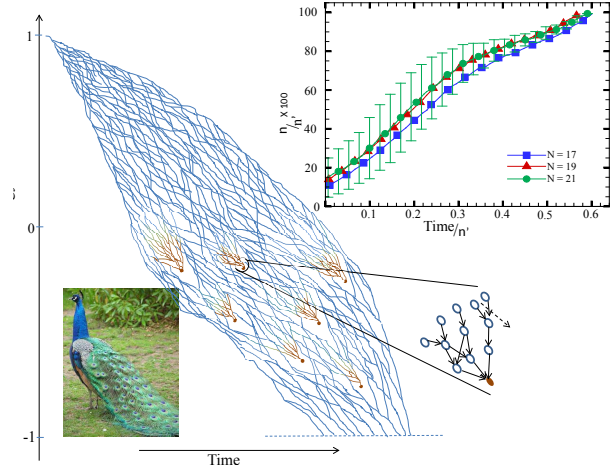}}
\caption{A schematic view of the network evolution (Peacock tail Pattern). Every point of the blue lines represent a specific topology of the network. The energy of the system decreases from top to bottom as the time increases from left to right. It is clearly obvious by following the blue lines that some topologies of the network are unable to evolve to any desired topology. The red points represent the topology of the jammed (Pseudo-stable) states, where the red lines leading to them are the pseudo paths before reaching the jammed states. The energy level equal to $1$ refers to an antagonistic state where the energy level equal to $-1$ refers to a balanced state. The top right panel shows the probability of going from one point to a point closer to a jammed state. Note that $n$ and $n'$ are the number of disallowed links to flip and the total number of links respectively.}
\label{fig2}
\end{figure}

By talking of systems in which their members influence each other,
the concept of collectivity comes in to play
\cite{bar1,rijt,Shirazi}. The collective behavior plays effective on
the evolution of such systems, in a sense that it provides
foundations for the creation of communities. However, this is not in
contradiction with the least energy principle, in fact it supports
it \cite{marvel}. To state clearer; the selected path in order to
lower the energy, is the will of the collective behavior in the
system. In systems which evolve regarding Heider balance
theory, when the energy obtains its lowest value (the balanced
states), the system would perform either a single or double community; where the former is called a paradise state and the latter is called a bipolar state. The paradise state refers to a
system that all of its members are friends with each other, while
the bipolar state refers to a system that consists of two main
groups where the members of each group are friends with each other
without being friends with any member of the other group. In these
networks, there are many paths that eventually lead to balanced
states. If a route other than that leading to a balanced state is
taken, one would end up in a state of local minimum energy named as
a jammed (pseudo stable) state, see \cite{Antal}. Talking of energy,
Marvel et al. \cite{marvel} showed that if the energy is descending
from $1$ to $-1$ these jammed states only occur between zero and
$-1$. However, there is a direct proportionality between the energy
value of the occurrence of the jammed state and the
number of communities in the network. They showed that; as the
energy value of the jammed state occurrence tends to zero, the
system becomes more complex due to the increase of the number of
communities. In this work we show that near the local and global
balanced states where communities start to perform, the participation of
individuals in a collective act increases.

After this introduction we proceed with our aims; our intention is
to emphasis the characteristics and features of jammed states. The
question to be answered is that although we know that the balanced
state implies a minimum stress condition and hence a minimum energy
state, why do the jammed states come in to play? do jammed
sates perform as a sudden event over the evolution of the system? or
they are special states at the end of a desired path. In this line
we need to understand the procedure in which the jammed states
emerge from other events, see also Zhihao \cite{Wu}. However to give
a flavour of how these jammed state looks, we have sketched Fig. 1.
The jammed states are shown as red points where one has been
magnified for providing a closer look. These jammed states are
points that one experiences while moving towards lower energy levels
from left to right, and are dead ends with no further
freedom to move on. To highlight this issue we change the color of lines from blue to
red in order to show the formation of pseudo paths towards these
jammed states. The top right panel in Fig. 1 is an indicator
providing information the probability of getting a step closer to a
jammed state. In the following, we use the inverse
participation ratio method (IPR) \cite{Jamali,IPR2016} to quantify
the amount of individuals' participation in forming a state. We take
the evolution of IPR as an indicator to help recognize the jammed
states in advance.

\section{The evolving network}
In the context of social network dynamics, everything lies
on the fact that tension must be reduced. This is in a sense that
every node (as a main actor) intends to reduce its tension with its
surroundings. As a matter of fact, the ambition to reduce the
tension justifies the root that the network selects as it evolves.
In this line, a good theory has been proposed by Heider named as the
balance theory \cite{heider}, where he studied the relation
between two persons and their attitude regarding an event. He referred
positive/ negative signs for the links between every two nodes.
Cartwright et al. \cite{car} took a further step and developed this
theory for social network applications. In this context the positive
sign would indicate for instance; friendship \cite{kuneg,esmail},
attractiveness \cite{doreian2}, profit \cite{guha}, tolerant
\cite{agui}, where the negative sign would indicate their opposite.
Nonetheless, it is the concept of these positive or negative links
that provide basis for understanding what is going on around us
\cite{zachary, moore}, see also refs \cite{marvel2,kuak} for
analytical and refs \cite{leskovec,szell,xiao,esmail} for numerical
models. In this model, the network is considered completely
connected, meaning that every node is related to all other nodes,
see e.g. refs \cite {James, Deok} for a mathematical description of
neighbour node effectiveness. This enables modeling the relations
between students in a classroom, members of a club
\cite{rijt,doreian}, representatives of countries in the United
Nations \cite{axel,moore} and epidemic spreading on evolving signed
networks \cite{epidemic}.

Now this is how it goes; the relation (link) between the members (nodes) $i$ and $j$ is represented by $S_{ij}$, where if $i$ and $j$ are equal, we would have $S_{ii}=0$. If the status between two nodes is friendship/enemy, we would have for $S_{ij}$ the values of 1 and -1, respectively \cite{car}. This imposes that for a three node system which shapes a triangle, a balanced triangle is formed only when the product of the values assigned for the links ($S_{ij}$) between every two nodes has a positive sign. Hence, we would have two balanced and two unbalanced states. Note that the tendency is to have a balanced triangle, which implies that as the network evolves, the unbalanced triangles eventually become balanced. However, in case of a weighted network where not all the nodes are connected to each other, Estrada \& Benzi provided a method for measuring how balanced the situation is \cite{Estrada}. Since the balanced and unbalanced states for the triangles are respectively referred to as the negative and positive energies, a Hamiltonian definition for these triangles is provided
\begin{eqnarray}
H=\frac{-1}{\binom{N}{3}}\sum S_{ij}S_{jk}S_{ki},
\label{e1}
\end{eqnarray}
where $N$ is the size of the network \cite{axel}. Note that the energy is obtained by subtracting the number of unbalanced triangles from the number of balanced triangles and normalizing to the total number of triangles. Hence, the Hamiltonian in Eq. (\ref{e1}) obtains values only between $-1$ and $+1$. By having in hand the Hamiltonian of a system together with the trend in which the energy is reducing, a model for its dynamics could be provided.

We start by considering a completely unbalanced network where all relations are initially poor. The reason for this selection is to foresee all the possible paths appearing in the evolution of the network right from the beginning. This is how it goes; randomly select a link and switch it to its opposite sign, if only the total energy of the network is reduced (even if a local energy in one triangle is increased), this selection is accepted. If flipping the link does not change the total energy, it is accepted by probability $0.5$. Otherwise the selected link is rejected and another link must be chosen \cite{antal}. If the process guides the system to its minimum global energy state or on other words a state that all triangles have become balanced (stable state); two conditions would be attained based on the initial conditions of the network. The system could either attain a state called "paradise" or a state called "bipolar". Now if the system does not finally become paradise or bipolar, it has surly been trapped somewhere, it is then that the system is in a jammed or a pseudo-stable state. Note that in jammed states although unbalanced triangles still exist, but any changes in the system would lead to an increase of energy.

\section{Results and discussions}
A community is not just constructed by its constituents but also by the relation between its constituents. It is due to this relation that macroscopic concepts like social and cultural laws in societies and magnetization in physics emerge. This not only affects the constituents individually, but also affects their collective behaviour, which puts the dynamics and evolution of their community in order. Although it is believed that the path towards the minimum energy state is always selected by the system, whether if there are specific patterns leading to the stable states? Now if there are specific paths, at what stage or time before reaching the stable states do they perform? But are all the paths towards the minimum energy state probable? or is there some desired paths? The results presented here prove that the collective behavior of the constituents play a significant role on the path of their evolution. Near the local and global minimum energy states some patterns start to form which host collective behaviors. In a sense that local minimum tension states in the middle of the network dynamics do not happen randomly. 

Before continuing any further we ought to define a new concept in this context named as participation, and state its difference with community. Community refers to nodes that have little separation lengths or are somehow consented with each other. But  participation refers to a condition where all the constituents have a mutual concern. Now this mutual concern on a matter could have various aspects due to the different viewpoints of the constituents. In case of a full agreement between the constituents, the concept of community comes into play, while in case of a complete disagreement, chaos arises, leaving a chaotic situation on our hands. However these are two limiting situations where depending on the importance of the issue, the system could be in any situation in between these two limits. Now being in either of the two limiting cases means that the issue under consideration is of high importance, which requires a high participation of the constituents.

\begin{figure}[!htb]
\centerline{\includegraphics[trim = 10mm 5mm 10mm 15mm, clip,width=.50\textwidth]{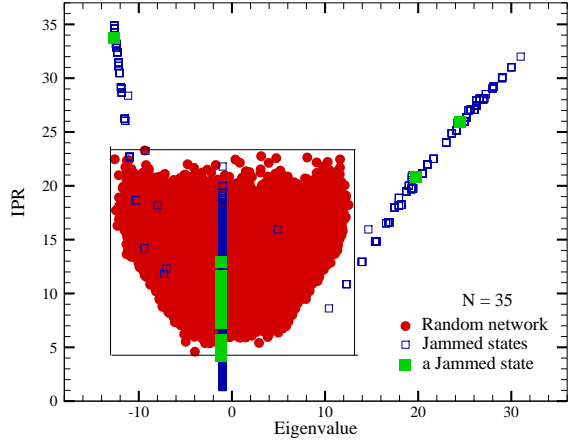}}
\caption{The inverse participation ratio (IPR) in terms of the eigenvalues for random (red circles) and jammed state (blue squares) networks; where the networks contain $35$ nodes. The green squares represent the eigenvalues of a specific jammed state. Note that three of these eigenvalues are located outside the bulk of the random network.}
\label{fig3*}
\end{figure}

\begin{figure*}
\centerline{\includegraphics[trim = 10mm 1mm 17mm 15mm, clip,width=.80\textwidth]{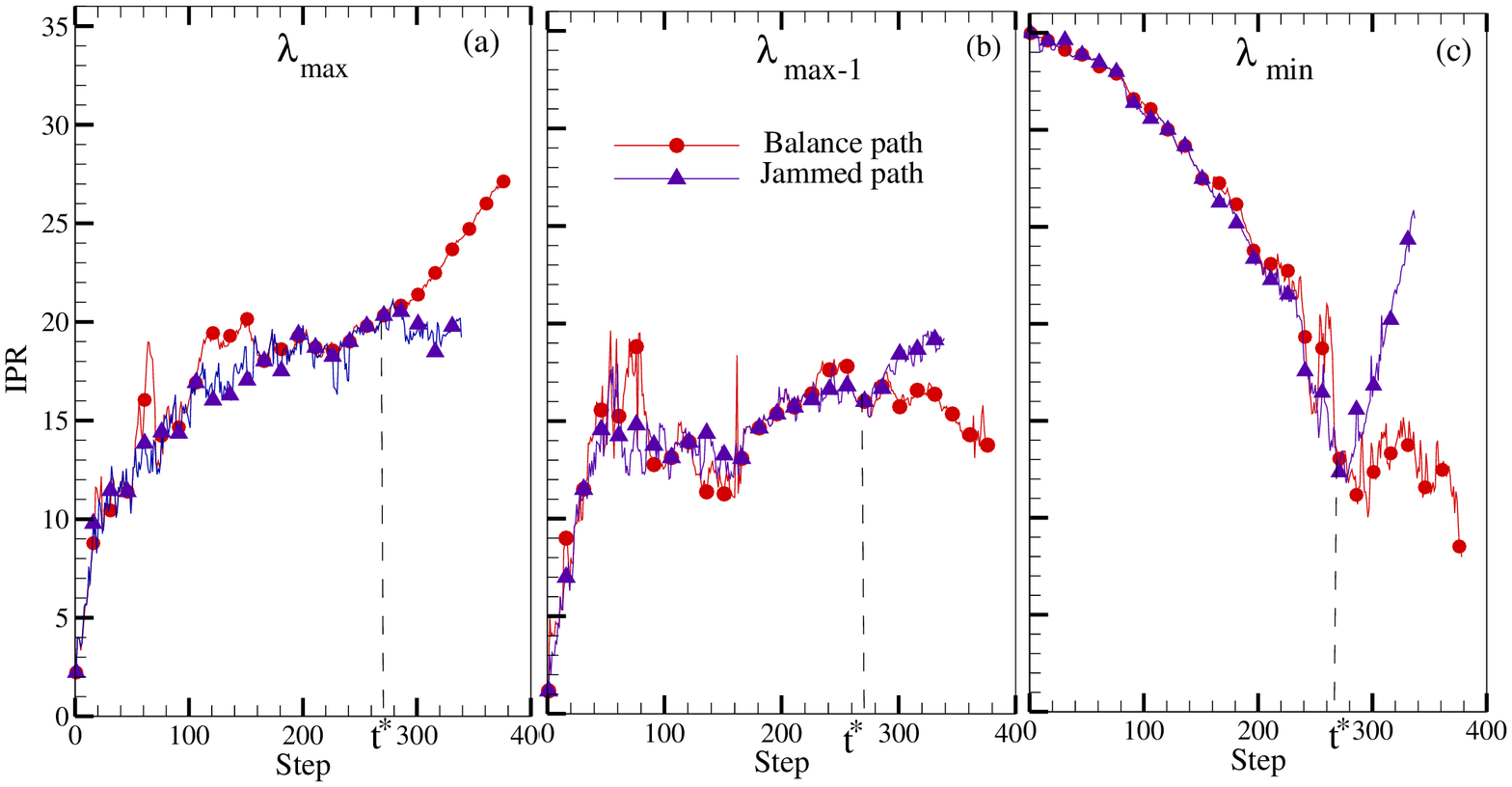}}
\caption{The evolution of the inverse participation ratio. The left, middle, and the right panels show the IPR evolution corresponding to the greatest, second greatest, and the smallest eigenvalues, respectively.}
\label{fig3}
\end{figure*}

Marvel et al. \cite{marvel} showed that when a system has more than two communities, the system is in a jammed state, where the number of communities is directly proportional to their energy levels. In other words, when the number of communities increases, the jammed states perform in higher energy levels. We find that if we are on a path that has a connection to a jammed state; the further away from the jammed state we are, the more chances we have to avoid ending up at the jammed state. This means that near these states the number of possible links which can be flipped and take the system in a path away from a jammed state decrease very fast. In fact, there is a characteristic length scale from a jammed state that greater than that, the chance of entering a root towards a jammed state is not provisioned, or in other words is random, see the panel inside Fig. 1. Regarding the new introduced concept, participation, we show that in the neighbourhood of these jammed states, the constituents participation start to increase, in a sense that the paths that would lead to the jammed states could be recognized in advance.
This clearly indicates that since jammed states are proofs for the existence of communities, the act of participation increases as the distance to the jammed state gets smaller than the characteristic length scale. This leads us to conclude that participation is directly proportional to the formation of communities. Strictly speaking, participation is a sign of community formation.

\subsection{Role of participation}
We now proceed by conducting a deeper exploration inside the network by shedding light on the organizers of all this. Thus, we study the jammed and balanced states regarding the behaviour of the eigenvalues, eigenvectors together with the participation ratio. Note that although the eigenvectors are usually used for community detection \cite{newman,newman2,kuneg2,samani,krz}, but here we use them as a measure for the participation. For instance, in a network consisting of two communities, it is claimed that the positive/negative components of the eigenvector corresponding to the greatest eigenvalue, is responsible for their detectance \cite{samani,capo}. But in a network consisting of several communities, one can not just rely on the information provided by the greatest eigenvalues \cite{samani,donetti}. It is in such cases that looking at more eigenvalues would provide more reliable information. Note that although it is the sign of the components of the eigenvector that indicates the community that hosts a node, but it is the magnitude of the components of the eigenvector that provides information on the amount of a node's participation. In other words; the greater the magnitude of the eigenvector components, the greater the share of participation. The question that arises here is on the value of this participation. The method implemented in the present study for estimating the amount of participation of each node is the inverse participation ration (IPR) obtained by \cite{slania,Namaki1,Jafari}
\begin{equation}
IPR=\frac{1}{\sum_{i=1}^{N}V_{i}^4},
\end{equation}
where $V_i$ is the $i$th component of the eigenvector corresponding to a specific eigenvalue. Note that the efficiency of the participation is directly proportional with the distance of $V_i$ from zero, no matter in what direction. This means that the state of minimum participation where all the components of the eigenvector are zero, except for one (which is equal to unity), the IPR is in its minimum state. Now if the other components of the eigenvector start to increase, the IPR would also increase until it attains its maximum value. To comply with the aims of the present work we need to understand the IPR variations of the eigenvalues that contain information.

Consider a network which is fully connected and undirected, the interaction matrix of its constituents is symmetric with all diagonal components equal to zero. Note that since the matrix is diagonal its eigenvalues are real. In this stage it is important to flag out the most effective eigenvalues, therefore we compare the eigenvalue spectra of the jammed states with the eigenvalue spectra extracted from a random network with positive/negative links, see Fig. 2. Due to the nature of all random processes, a random network hardly possesses any information to rely on. Based on this statement, one could simply deduce that any deviation from a random curve would increase the reliability of the information extracted from it \cite{Namaki2}. In other words, the reliability of the results extracted from a curve is directly proportional to the deviation from its corresponding random curve. In Fig. 2, the divergence of the eigenvalues of the jammed states from the random network (bulk) in terms of the IPR is shown for a $35$ node network. The red circles represent the eigenvalues of a random network ensemble, the blue squares represent the eigenvalues of various jammed states, and the green squares represent eigenvalues for a specific jammed state. It could readily be noticed that two of the eigenvalues indicated by green squares lie at the right of the bulk and one lies at the left of the bulk. As stated earlier, the eigenvalues located inside the bulk are not much of importance. This is due to their overlapping with other eigenvalues inside the bulk corresponding to a random network. But the important thing is the eigenvalues outside the bulk, where the green eigenvalue on the left of the bulk represents the lowest eigenvalue $\lambda_{min}$ of the jammed state, and the two on the right represent the greatest $\lambda_{max}$ and the second greatest $\lambda_{max-1}$ eigenvalues of that jammed state.

We draw attention to the informational content of the eigenvalues and study the IPR evolution of the states in our working network. Note that in this line we need only focus on three eigenvalues (on their ways to the balanced or jammed states) which lie outside the bulk of the random network, see Fig. 2. We investigate the evolution of IPR corresponding to these three eigenvalues.
By having in hand the information extracted from these eigenvalues, it will be possible to understand the IPR evolution for that network. It enables estimating the participation percentage of the nodes. This enables the prediction of the pattern of the network's dynamics that leads to jammed states. Fig. 3 illustrates the evolution of IPR for the three working eigenvalues of our $35$ node network. The IPR corresponding to the eigenvector of the greatest eigenvalue, $\lambda_{max}$, has an elevating trend towards the stable state no matter balanced or jammed. In such a situation the stable state is achieved by participation at its highest level. But the IPR that corresponds to the eigenvectors of the other two eigenvalues, $\lambda_{min}$ and $\lambda_{max-1}$, take their roots according to their tendency towards either balanced or jammed states. Hence, the information of the paths leading to the jammed states are mostly possessed by $\lambda_{min}$ and $\lambda_{max-1}$, see the two right panels of Fig. 3 which readily illustrates their divergence.


It is now time to illustrate two examples for better understanding the statements of the previous paragraphs. The first example is in application to societies, while the second deals with magnetic media. For the former example, consider a society that every person constituting it has its own intentions. When the dynamics of such a society tends towards a decreasing tension level, the people would start collaborating with each other due to the peace in mind that the low tension level has offered them. In such a system the people are likely to talk to each other more, this makes people notice their common intentions. The result would increase the participation rate of the people in the society. Due to the common intentions of the individuals in that society the process of adding up occurs, which would eventually lead the society to be in either a balanced or jammed state. Now if the adding up of common people with common interests results in the creation of one or two groups, the state is considered balanced. But if more than two groups of people are created, the society is considered to be in a jammed state. For the latter example consider a magnetic system; reduce the heat and obtain a lower energy level, the result would be the performing of various magnetic domains. All magnetic field lines in one domain point at the same direction, but each domain has its own direction. Now if there exists one or two domains with different directions, the state is balanced. But if more than two domains with different directions exist, the state is jammed.

\subsection{A qualitative approach}
By putting all the pieces together we can now state that for a network it is not simple to predict what happens next. There seems to be a secret in the dynamics of the networks that gradually decreases the number of options available for avoiding a jammed state. To be precise, in a network there is a potential that bounds us in a path that makes it inevitable to end up in a specific jammed state. This means that the chance for taking paths that lead to lower energy states rapidly fade away. This statement could also be supported by a qualitative approach in application to dynamics of the links. Our working links are the unbalanced links defined by $S_{ij} \sum_{k \neq i,j} S_{ik} S_{kj} < 0$, where their sign flip directs the evolution process. Nonetheless, a matrix representation of the set of unbalanced links stands only if the constrain $\lbrace \mathrm{sgn}((S)_{ij}) \neq \mathrm{sgn}((S^2)_{ij}) \rbrace$ is respected. Now a flip of an unbalanced link would affect the next system of unbalanced links in two manners: in one way it changes the sign of $S_{ij}$, and in the other it changes all the entities of the $i$th row and $j$th column of $S^2$ by $\pm 2$. Thus, the evolution process of the system could be influenced by $S^2$ only if their values are near zero. Otherwise, the only term that contributes to the evolution process is $S$.
Now if the $S^2$ entities are far from zero, the evolution process would send out one member of the unbalanced set in the course of every iteration. In Fig. 4 a validation of the assertion is provided by demonstrating the fraction of the unbalanced links for each time step for both dead ended balanced and jammed states. Note that the plot in Fig. 4 is retrospective, starting from the last time step. The linear behavior of the curves just approaching the final states is due to the fact that the absolute values of $S^2$ entities could possibly be much greater than zero. Therefore it supports the statement that flipping a link removes only one link from the set. Such a linear behavior implies that the system is trapped in an attractor basin a few steps left to the fixed point. This provides basis for developing a criterion to enable forecasting the system. This clearly proves the constriction possessed by the phase space in the vicinity of the final states.

\begin{figure}[t]
\centerline{\includegraphics[trim = 8mm 1mm 19mm 14mm, clip,width=0.5\textwidth]{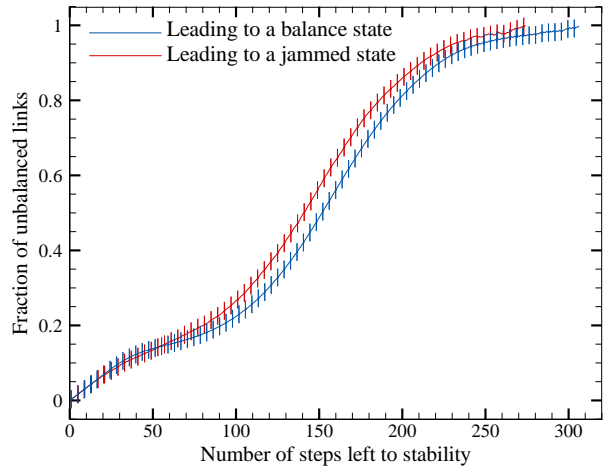}}
\caption{Dependence of the Fraction of unbalanced links on the time steps left to balanced (blue curves) and jammed states (red curves). Note that the plot is retrospective.}
\label{fig4}
\end{figure}

\section{Conclusions}

In a network with a high level of complexity, the balanced theory
implies the type of interactions between the constituents of the
network to provide prohibition on some paths leading to the minimum
energy state.
Out of all the existing possible ways for the evolution of a network,
sometimes there are only few alternative paths of change. However, in some cases there isn't even another. It seems while evolving, there is a push to lead the network to special points.
The secret that has been unveiled in the present work is nothing but participation. As a
matter of fact it is the participation of the constituents of a
society that dictates the paths towards the lower energy levels.
%
The roots towards the minimum energy levels, rapidly
decline, which would eventually lead you to a path with no return.
It is in such a situation that you are almost at a local energy
level or a jammed state. Hence the accumulation or adding up of
constituents on jammed state is a product of participation in the network.
When a system enters a path towards a jammed state, it is condemned to limbo where any effort could not change the destiny of the system. We consider the inverse participation ratio method (IPR) as a suitable indicator that can quantify the participation in forming a specific state. Here we investigated the evolution of IPR corresponding to the eigenvectors of eigenvalues containing information.

Provided that the picture illustrated in the present study is well understood; a sophisticated program based on the architecture of the community resulting in individuals best interest could be directed and managed. Also by learning the prospective paths towards glory, a lovely community could be developed.\\\\


Acknowledgment: The authors would like to thank Prof. Muhamad Sahimi for his constrictive comments and helping to edit the manuscript.
The research of GRJ was supported by the Higher Education Support Program of OSF and the Central European University.

\end{document}